\documentclass[twocolumn,superscriptaddress]{revtex4}
\usepackage{amsmath, amsfonts, amssymb, graphicx, color,tabularx}

\newcommand{\lastequal}{Corresponding authors. These authors contributed equally.}

\begin{document}

\title{NoisET: Noise learning and Expansion detection\\
  of T-cell receptors}

\author{Meriem Bensouda Koraichi}
\affiliation{Laboratoire de physique de l'\'Ecole normale sup\'erieure,
  CNRS, PSL University, Sorbonne Universit\'e, and Universit\'e de
  Paris, 75005 Paris, France}
\author{Maximilian Puelma Touzel}
\affiliation{MILA, University of Montreal, Montreal, Canada}
\author{Andrea Mazzolini}
\affiliation{Laboratoire de physique de l'\'Ecole normale sup\'erieure,
  CNRS, PSL University, Sorbonne Universit\'e, and Universit\'e de
  Paris, 75005 Paris, France}
\author{Thierry Mora}
\thanks{\lastequal}
\affiliation{Laboratoire de physique de l'\'Ecole normale sup\'erieure,
  CNRS, PSL University, Sorbonne Universit\'e, and Universit\'e de
  Paris, 75005 Paris, France}
\author{Aleksandra M. Walczak}
\thanks{\lastequal}
\affiliation{Laboratoire de physique de l'\'Ecole normale sup\'erieure,
  CNRS, PSL University, Sorbonne Universit\'e, and Universit\'e de
  Paris, 75005 Paris, France}

\begin{abstract}
 % abstract
 High-throughput sequencing of T- and B-cell receptors makes it possible to track immune repertoires across time, in different tissues, in acute and chronic diseases and in healthy individuals. 
However quantitative comparison between repertoires is confounded by variability in the read count of each receptor clonotype due to sampling, library preparation, and expression noise.
We review methods for accounting for both biological and experimental noise and present an easy-to-use python package \textit{NoisET} that implements and generalizes a previously developed Bayesian method. It can be used to learn experimental noise models for repertoire sequencing from replicates, and to detect responding clones following a stimulus. We test the package on different repertoire sequencing technologies and datasets. We review how such approaches have been used to identify responding clonotypes in vaccination and disease data. \\
Availability: \textit{NoisET} is freely available to use with source code at \url{github.com/statbiophys/NoisET}.

\end{abstract}

\maketitle
 % maintext
\section{Introduction}

Cells of the adaptive immune system, T and B lymphocytes, recognize molecules foreign to our body and protect us against pathogenic threats. These cells also have the ability to eliminate cells that harbor anomalies, such as cancer cells. Lymphocytes perform this discrimination task between potentially dangerous and normally functioning ``self" molecules using specialized receptors on their surface that constantly sample and bind molecules in our organisms. Each cell has one type of receptor and the system relies on a large diversity of a repertoire of different receptors expressed on over $10^9$ different B or T cells to protect the organism against infections~\cite{Robins2009, MoraWalczak2016, Lythe2016, DeWitt2016}. 

The composition of the repertoire contains information about past infections and conditions. Reading this information requires quantitatively understanding the natural repertoire dynamics. Upon recognition of a pathogenic molecule, the recognizing cell proliferates making many cells with the same receptor, forming a clone, which enables fast infection clearance. New cells are constantly produced and introduced into this diverse repertoire. Additionally to specific stimulation, cells also undergo random divisions. Each cell has a finite lifetime and clones can go extinct if all the cells of that clone die. Together these processes define a natural dynamics of the repertoire, which leads to a constantly changing set of different cells present at different frequencies. 

High-Throughput Repertoire Sequencing (RepSeq) of T and B cell receptors (TCR and BCR)  \cite{Weinstein2009,Robins2009,Boyd2009a, Benichou2012, Six2013,Robins2013a,Georgiou2014,Heather2017,Minervina2019, NatureAIRR} enables us to study the dynamics of lymphocytes at the resolution of single clones, by comparing their concentrations across timepoints or conditions. To detect biologically relevant clones, one must be able to distinguish true differences in clone frequencies from experimental noise. This variability has three sources. First, laboratories use various sequencing  and sample preparation protocols using either gDNA or cDNA (with or without unique molecular identifiers), with different outcomes in terms of amplification bias and errors \cite{Heather2017, Barennes2020}.  This makes it difficult to reliably estimate TCR or BCR clonal frequencies from sequence counts. Secondly, in the case of cDNA based sequencing, these uncertainties are not solely due to different sample preparation but have a more fundamental, biological source. mRNA is produced in bursts \cite{Elowitz2002,vanOudenaarden2002,Xie2006, tanaguchi_xiescience, HornosPRE}, which adds a natural longtailed noise to the sequencing read distribution.  Thirdly, one must translate immune information contained in a few milliliters of blood to the whole repertoire. To describe these sources of variability, one needs a probabilistic approach.

Puelma Touzel et al.~\cite{PuelmaTouzel2020} developed a 
statistical model to identify responding clones using sequence counts in longitudinal RepSeq data. This model captures features of a repertoire response to a single, strong perturbation (e.g. yellow fever vaccination), giving rise to a fast transient response dynamics. The method was proposed as an alternative to commonly used tests such as Fisher's exact test \cite{Balachandran2017} or beta binomial models \cite{Rytlewski2019}. Its main innovation is to account for the different sources of biological and experimental noise in the clone count measurements in a Bayesian way, allowing for a more reliable detection of expanded or contracted clones.

Here we briefly review the ideas behind the method that calibrates the noise and we  introduce NoisET (Noise sampling learning and Expansion detection of TCRs), an easy-to-use python package that implements this method and extends it to datasets of diverse origin describing the clonal repertoire response to acute infections. We also review several applications of this approach.

\section{Model}

In order to correctly identify expanding or contracting clonotypes, whether after direct antigenic stimulation or due to random cell division and death, we need to correctly separate biological and experimental noise from the lymphocyte dynamics. The main idea behind the Bayesian probabilistic modeling method implemented in the NoisET software is learning probabilistic distributions describing sampling and experimental noise from empirical frequencies of TCR counts in biological replicate samples from the same individual.  In this section we introduce the two types of models implemented in NoisET: the noise model and the response model. 

\subsection{Modeling experimental noise}~\label{Methods_noise}
 TCR sequencing (TCRseq) methods, depending if they are based on DNA or RNA input,   produce data with different characteristics. For example, RNA-based methods  allow for the usage of unique molecular identifiers (UMI) to limit PCR amplification bias and sequencing errors. Non-UMI methods are better in capturing rare clones which motivates their frequent usage \cite{Barennes2020}. During this first step, NoisET learns at the same time the exponent  of the underlying power-law TCR frequency distribution, $\rho(f)$~\cite{Weinstein2009} and the parameters of error model between the empirical abundance of one specific TCR clone $\hat{n}$ and its true frequency $f$: $P(\hat{n}|f)$. NoisET has also the power to learn these distributions constraining the size of the clones we want to take into account for the analysis. 
 
For each TCR clone,  the likelihood to sample $\hat{n}$ reads from the first biological replicate and $\hat{n}'$ reads from the second biological replicate is:
\begin{equation} \label{noise-likelihood}
\mathbb{P}(\hat{n}, \hat{n}' | \Theta) = \int_{f_{{\rm{min}}}}^1 df \rho(f| \Theta) P(\hat{n}|f, \Theta) P(\hat{n}'|f, \Theta),
\end{equation}
where $\Theta $ are the parameters of the noise model which define the error model $P(\hat{n}|f)$, 
 $f_{{\rm{min}}}$ corresponds to the minimum clonal frequency for each individual, and $\rho(f)$, which is the clonal frequency prior known to be a power-law distribution $\propto f^{\alpha}$ ~\cite{Weinstein2009, MoraWalczak2019}. 
NoisET learns the parameters of the noise model $\Theta$ by maximizing the log-likelihood of the observed TCR counts, $\hat{n}, \hat{n}'$, from the two biological replicates:
\begin{equation} \label{eq:noise}
\begin{aligned}[t]
  \Theta^*  &=  \underset{ \Theta }{\text{argmax}} \  \displaystyle \prod_{i =1}^{N_{obs}} \mathbb{P}(\hat{n}_i, \hat{n}'_i | \Theta) \\
\end{aligned}.
\end{equation}

Since in RepSeq samples we only partially sample an individual's repertoire, the likelihood in Eq.~\ref{noise-likelihood} needs to be modified. We condition the likelihood on observing that specific clone in at least one of the two replicates: $\mathbb{P}(\hat{n} + \hat{n}' > 0)$. The modified likelihood becomes  $ \mathbb{P}(\hat{n}, \hat{n}' | \hat{n} + \hat{n}' > 0 ) =  \mathbb{P}(\hat{n}, \hat{n}', \hat{n} + \hat{n}' > 0 ) / \mathbb{P}(\hat{n} + \hat{n}' > 0) $. 
We can also choose to learn the noise model only on clones having a size larger than a certain threshold. In this case the likelihood in Eq.~\ref{noise-likelihood} becomes: $\mathbb{P}(\hat{n},\hat{n}' | \hat{n}> \hat{n}_{th} , \hat{n}’> \hat{n}_{th} ) = \mathbb{P}(\hat{n}, \hat{n}', \hat{n}> \hat{n}_{th}, \hat{n}’> \hat{n}_{th}) / \mathbb{P}(\hat{n}> \hat{n}_{th}, \hat{n}’> \hat{n}_{th})$. \\

To take into account different possible sources of noise due to the various RepSeq method, NoisET gives the choice of three different probabilistic distributions to learn the biological and experimental noise in the measured TCR abundances, $P(\hat{n}|f)$:

\begin{itemize}
\item The Poisson distribution, $P(\hat{n}|f) = {\rm{Poisson}} (f N_r)$. In this case, the noise parameters $ \Theta$ are the exponent $\alpha$ of the clone-size distribution $\rho(f) = C f^{\alpha}$ and  the minimum clonal frequency in Eq.~\ref{noise-likelihood}, $f_{\rm min}$. $N_r$ is the total number of reads in the sample.

\item The negative binomial distribution: $P(\hat{n}| f)=\mathrm{NegBin}(\hat n ;N_rf,N_rf+a(N_r f)^b)$, where $\mathrm{NegBin}(n;x,\sigma)$ is a negative binomial of mean $x$ and variance $\sigma$. In this case, $ \Theta=(\alpha , a, b, f_{\rm{min}})$ with $\alpha$, $f_{\rm{min}}$ being the same parameters as described for the Poisson distribution, and $a$ and $b$, the parameters of the negative binomial distribution. 

\item The negative binomial combined with a Poisson distribution: $P(\hat{n}|f)  = \displaystyle \sum_{m_i}^{\infty} P(\hat{n}|m_i) P(m_i|f)$, 
with $P(m_i|f)  = \mathrm{NegBin}(m_i ;fM,fM+a(fM)^b)$ and  $P(\hat{n}|m_i) = {\rm{Poisson}} (m_i N_r/M)$. A clone of size $f$ appears in a sample containing $M$ T-cells on average as $fM$ cells. To account for over dispersion, the number of cells associated to a specific clone is $m$ and follows a negative binomial of mean $fM$ and variance $fM+a(fM)^b$. For each clone the empirical abundance read in the biological sample is distributed according to a Poisson distribution with mean $m N_r / M$. For this model $ \Theta=(\alpha , a, b, M, f_{\rm{min}})$. 
\end{itemize}

While the mathematical framework is the same, when applied to identifying expanding clonotypes NoisET uses noise parameters inferred at both time points, contrary to the approach taken in \cite{Pogorelyy2018} and \cite{PuelmaTouzel2020}.  Experimental conditions at both time points can vary and it is important to use both sets of parameters $\Theta$, to have the correct form of $P(\hat{n}|f, t_1)$ and  $P(\hat{n}|f, t_2)$. The exponent of the power-law $\alpha$ and $f_{\rm min}$  in  Eq.~\ref{noise-likelihood} are the learnt values inferred at the time point for which sequencing depth is the larger. 

\begin{figure*}
\includegraphics[width = \linewidth]{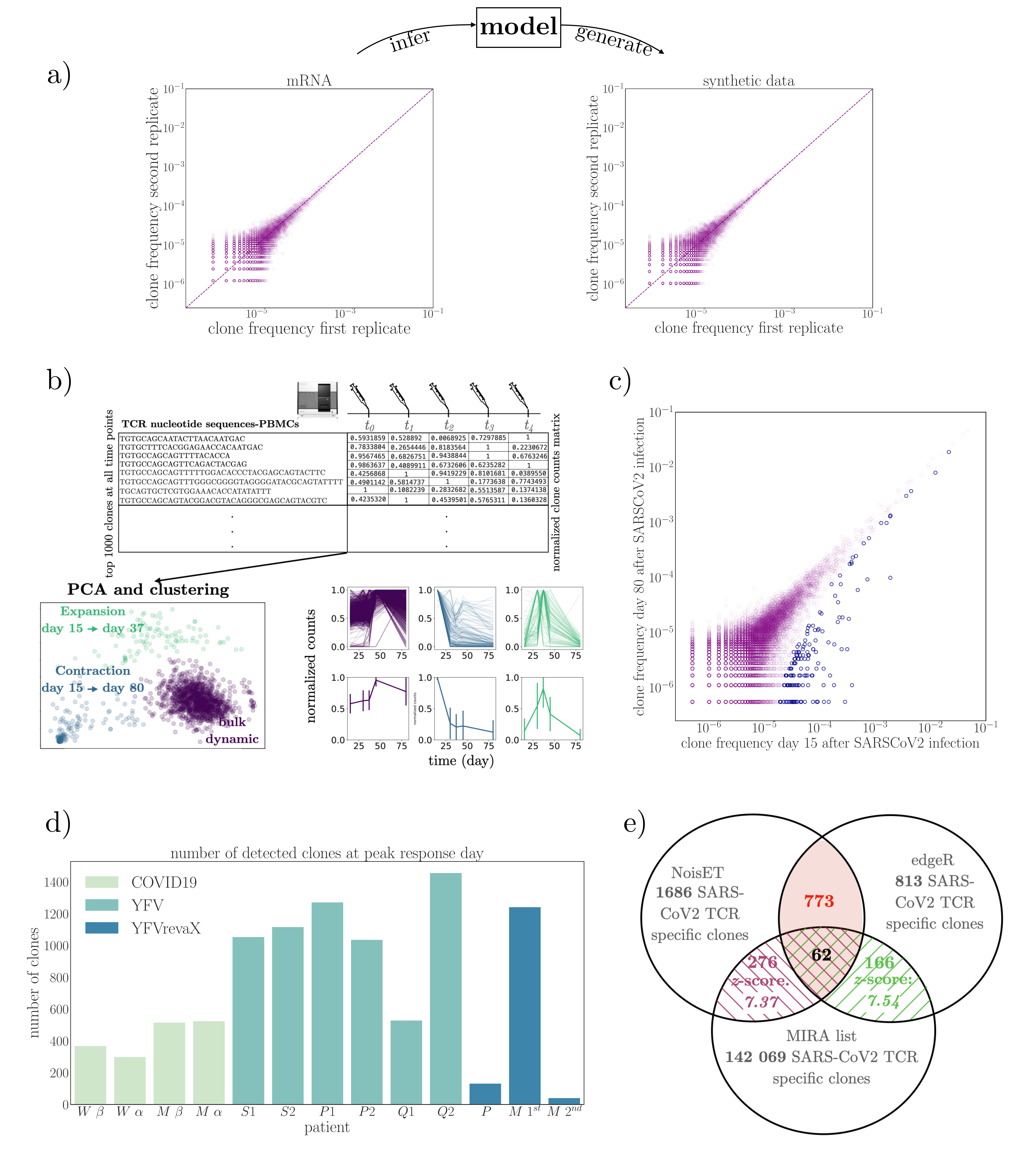}
\caption{{\bf(a)}  Scatter plots of sequence counts from two biological replicates from \cite{Pogorelyy2018} (left). NoisET learns a statistical model of sequence frequencies and observed counts from these data (here with negative binomial sampling noise model), which can then be used to generate realistic synthetic data (right). {\bf (b)} 
PCA (Principal Component Analysis) performed on the matrix composed of the normalized clone counts of the top 1000 clones present at every time point of the longitudinal dataset. The clustering of the data projected on the two first principal components enables us to understand different kinds of dynamics for three clusters of clones here (bottom left). The number of clusters can be adjusted in NoisET and should be tested. In this  example, this pre-analysis of the longitudinal dataset enabled us to find a significant contracting dynamical trend between day 15 to day 85 and a significant expansion trend between day 15 and day 37 following a  mild COVID-19 infection \cite{Minervina2021} (bottom left). The top plots show the individual trajectories in each trend, the bottom plots the average with standard variation error bars. 
{\bf (c)} Scatter plot of contracted clones from day 15 to day 85 after a mild COVID-19 infection \cite{Minervina2021}. Clones detected as contracting by { NoisET} are shown in blue. {\bf (d)} The number of responding clones detected by NoisET  (using a two step noise model) for 3 studies: donors M and W (with both $\alpha$ and $\beta$ TCR chains) in response to a SARS-CoV-2 infection between days 15 and 85 post infection \cite{Minervina2021}; 6 twin donors (S1 through Q2, only $\beta$ chain) between days 0 and 15 following yellow-fever vaccination \cite{Pogorelyy2018}; and yellow-fever first (M) and second vaccination (M and P) \cite{Minervina2020}. {\bf (e)} Venn diagram showing the overlap between the number of called responding TCR clones by both NoisET and edgeR after a mild COVID-19 infection \cite{Minervina2021}. The  $z-$scores and  $p-$values of the common clones found by both methods and the MIRA data base. Plots {\bf a-c} are standard NoisET output.
}\label{fig:01}
\end{figure*}

\newpage

\subsection{Detecting responding TCR clones}~\label{Methods_response}

To account for differential expression after antigenic stimulation, NoisET implements the approach of previous work~\cite{Pogorelyy2018, PuelmaTouzel2020} that introduced a selection factor $s$ defined as the log-fold change between a clone's frequency at time $t_1$ $f(t_1) = f$, and the frequency at time $t_2$, $f(t_2) = fe^s$.  A prior is assumed over the variable $s$,  $P(s| \gamma, \bar{s}) = \gamma \exp(- |s|/ \bar{s})/(2\bar{s}) + (1-\gamma)\delta(s)$, with $0 \leq \gamma \leq 1$ the fraction of responding clones and $\bar{s} > 0$, their typical effect size. The likelihood associated to observing a clone with empirical abundances $\hat{n}_1$ at time $t_1$ and $\hat{n}_2$ at time $t_2$ integrating the prior knowledge over the log-fold change $s$ is the following:

\begin{equation} 
\begin{split}
 & \mathbb{P} (\hat n_i(t_1)=\hat n_1,\hat n_i(t_2)=\hat n_2 )| \gamma, \bar{s} )= \\
 & \int \int {df_1} \rho(f_1) ds P(s| \gamma, \bar{s})  P(\hat n_1|f_1) P(\hat n_2|f_1e^s).
\end{split}
\end{equation}

The parameters $(\gamma, \bar{s})$, are learned by maximizing the likelihood of the count pair data taken at two given time points:
\begin{equation}\label{eq:inf}
\left({\gamma^*} ,{\bar{s}^*}\right)   =  \underset{ \left( {\gamma} , \bar{s}  \right)}{\text{argmax}} \  \displaystyle \prod_{i =1}^{N_{obs}} \frac{\mathbb{P} (\hat n_i(t_1),\hat n_i(t_2) | \gamma, \bar{s}, \Theta(t_1), \Theta(t_2) )} {\mathcal{Z}(\gamma, \bar{s})}, 
\end{equation}
with $\mathcal{Z}(\gamma, \bar{s})$ a normalization factor accounting for the probability to observe TCR clone counts in both analyzed samples and $\Theta(t_1), \Theta(t_2)$ the noise parameters learned at both time points $t_1$ and $t_2$ with NoisET. These two parameters were then used to compute the posterior $\mathbb{P}(s|\hat n_i(t_1),\hat n_i(t_2))$: 
\begin{equation}\label{eq:post}
 \mathbb{P}(s|\hat n_1,\hat n_2)=  \frac{ \mathbb{P}(\hat n_1,\hat n_2 | s, \gamma, \bar{s}) P(s| \gamma, \bar{s})}{\mathbb{P}(\hat n_1,\hat n_2))}.
\end{equation}

The knowledge of the log-fold change posterior (\ref{eq:post}) is used to discriminate expanded or contracted clones from the bulk between $t_1$ and $t_2$. In analogy with $p$-values, we define $p = \mathbb{P}(s \leq 0 | \hat n_1,\hat n_2, \gamma, \bar{s}, \Theta(t_1), \Theta(t_2))$, the probability corresponding to the null hypothesis of no expansion. If $p < \rm{threshold}$, the clone is classified as expanded. When looking at contraction, we use the same method reversing times $t_1$ and $t_2$ and looking at significant expansions from $t_2$ to $t_1$. The value of the threshold can be chosen by the user. In all the results presented in this review, the threshold was set to $0.05$, however no threshold was applied when identifying contracting clones. Another threshold on the median of the $\mathbb{P}(s|\hat n_1,\hat n_2)$ distribution can be applied to select for clones that are greatly expanded. 

The output of NoisET detection of responding clones is the list of statistical properties of the true log-fold change variable $s$  called according to the posterior  $P(s| \hat{n}_1, t_1, \hat{n}_2, t_2)$ learned from data after learning the noise and differential model (Eq.~\ref{eq:noise},\ref{eq:inf}). These statistics are mathematically defined in Table~\ref{tab:table-1} and are the values of $s$ that defines the first quantile $s_{1, \rm low}$, the median of the posterior $ s_{2, \rm  med}$, the value of $s$ that defines the third quantile $s_{3, \rm high}$, the mode of the posterior $s_{\rm max}$, the average of the posterior $\bar{s}$ and and the $p$-value like value defined as $P(s \leq 0 | \hat{n}_1, \hat{n}_2).$ 

\begin{table}
\begin{tabularx}{\linewidth}{ 
 | >{\raggedright\arraybackslash}X 
 | >{\centering\arraybackslash}X 
| >{\raggedleft\arraybackslash}X | }
\hline
Feature & Description  \\
\hline
$s_{1, \rm low}$  & $\int_{-\infty}^{s_{1, \rm low}} \mathbb{P}(s|\hat n_1,\hat n_2) ds = 0.025$  \\
\hline
$ s_{2, \rm  med}$  & $\int_{-\infty}^{s_{2, \rm med}} \mathbb{P}(s|\hat n_1,\hat n_2) ds = 0.5$    \\
\hline
$s_{3, \rm high}$   & $\int_{-\infty}^{s_{3, \rm high}} \mathbb{P}(s|\hat n_1,\hat n_2) ds = 0.975$    \\
\hline
$s_{\rm max}$   & $\underset{ \left( s  \right)}{\text{argmax}} \mathbb{P}(s|\hat n_1,\hat n_2) $      \\
\hline
$\bar{s}$   &   $\int_{-\infty}^{+ \infty} s \mathbb{P}(s|\hat n_1,\hat n_2) ds $   \\
\hline
$ 1 - \mathbb{P}(s >0)$   &    $ \int_{-\infty}^{+ \infty} s \mathbb{P}(s|\hat n_1,\hat n_2) ds $\\
\hline
\end{tabularx}
\caption{\label{tab:table-1} Mathematical definition of statistical properties of the hidden variable $s$, the log-fold change of counts of a given clone, computed from the posterior distribution $P(s| \hat{n}_1, t_1, \hat{n}_2, t_2)$, learned from the noise and differential model. The output of NoisET when detecting significantly expanded clones consists of the list of clones that are detected to have respectively increased or decreased in term of abundance associated with these specific $s$ characteristics. }
\end{table}

\section{Features}

NoisET has two main functions: (1) inference of a statistical null model of sequence counts and variability, using replicate RepSeq experiments, as described by the models presented in section~\ref{Methods_noise}; (2) detection of responding clones to a stimulus by comparison of two repertoires taken at two timepoints, as described by the models presented in section~\ref{Methods_noise}. The second function requires a noise model, which is given as an output of the first function. Both functions require two lists of sequence counts associated to each TCR or BCR present in the repertoires: from replicate experiments for the first function (Fig.~1a left), and from repertoires before and after the stimulus for the second function (Fig.~1a right). In addition, NoisET has features for detecting the time points to be compared, to simulate natural immune repertoire dynamics, and to estimate diversity. 

All functions are described in a README and notebooks available on the Github repository (\url{https://github.com/statbiophys/NoisET}). A tutorial explains the different functions of NoisET.

\subsection{Detecting the peak moment of the response}

When more than two time points are available, and when the timescales of the dynamical response of the TCR repertoire to an acute infection are not known, it is difficult to know which pairs of time points in longitudinal data can be informative about responding clonotypes. A method based on Principal Component Analysis (PCA) of longitudinal trajectories was first used in~\cite{Minervina2020, Minervina2021} to identify the peak of the response (Fig ~1b). 
It uses the first two PCA components of the 1000 most abundant TCR clonotype frequencies normalized by their maximum post-infection values. The clustered trajectories identify different modes of  clonal abundance dynamics. NoisET includes a feature for performing this PCA on trajectories as a preliminary step to pick the best timepoints.

\subsection{Learning the noise model}

When learning a noise model from replicates, the user must pick the type of noise model, which describes how the sequence count in the RepSeq sample depends probabilistically on its true frequency in the blood. Choices are: a Poisson distribution, a negative binomial distribution, or a two-step model \cite{PuelmaTouzel2020}.
Once the parameters have been learned (Maximum Likelihood Estimation optimization algorithm), a generation tool can be applied to qualitatively check the agreement between data and model for replicates (Fig ~1a). We also successfully learned a null model from gDNA data \cite{Rytlewski2019}, which is included in the package example notebook.

\subsection{Detecting responding clones}

To use the second function to detect responding clonotypes, the user provides, in addition to the two datasets to be compared, two sets of experimental noise parameters learned at both times using the first function.
When replicates are not available for each time point or donor,
a common null model may be used for both timepoints. This should be done with caution, since even if both samples are produced with the same technology for the same donor, the sequencing depth and distribution of clone frequencies may vary between timepoints.
Finally the user provides two thresholds:
one for the posterior probability above which a clone is labeled as responding, and one for the median log-fold frequency difference above which detection is allowed.
The output is a CSV file containing a table of putative responding clones. The result is illustrated in Fig.~1c, which shows contracted clones (purple points) detected from day 15 to day 85 from a mild COVID-19 infection \cite{Minervina2021}.

Compared with software introduced in Ref.~\cite{PuelmaTouzel2020}, NoisET allows for conditioning on TCR clones sizes in the analysis, and for using a Poisson or negative binomial distribution for the experimental noise model.

\subsection{Generating trajectories}~\label{feat_gen}
Using NoisET, one can also generate \textit{in-silico} RepSeq samples, and their neutral dynamics following the stochastic population dynamics developed in \cite{Desponds2016}, and in \cite{BensoudaKoraichi2022}. The function takes as input the noise model method (negative binomial or Poisson), the noise model parameters at both time points, the number of reads at both time points, the duration of the simulations, and the values of $\tau$ and $\theta$ describing the global stochastic population dynamics. The neutral dynamics for each clone is defined by $ \frac{dn}{dt} = \left[- \frac{1}{\tau} + \frac{1}{2\theta} + \frac{1}{\sqrt{\theta}}{\eta(t)}\right] n(t)$,  with $n(t)$ -- the true somatic abundance for a clone belonging to an individual repertoire.

\subsection{Diversity estimator}
Learning the noise model is also helpful for computing diversity estimates which are known to be sensitive to sampling noise~\cite{MoraWalczak2019}. 
NoisET includes a diversity estimator $D_0 = N_{\rm obs}/(1-P(\hat{n}=0, \hat{n}'=0))$, with $N_{\rm obs}$ the number of clones observed in both replicates used to learn the experimental noise, and  $P(\hat{n}=0, \hat{n}'=0)$ the learned fraction of non-sampled clones from the repertoire. This value is expected to be close to 1. Evidently, the larger $N_{\rm obs}$, the deeper the sequencing is and so the diversity estimate is expected to be more trustworthy, assuming comparable quality of data generation.

\section{Applications of NoisET}

The method on which NoisET is based has been applied in two  published studies identifying clones involved in yellow fever vaccination~\cite{Pogorelyy2018} and SARS-CoV2 responses~\cite{Minervina2021}. In both cases, the analysis was performed on longitudinal TCR RepSeq cDNA data sets and from several different time-points, we were able to identify the peak of the response (expansion or contraction) thanks to the trajectory PCA method~\cite{Minervina2020} now encoded in NoisET. Fig.~1d reports the number of responding clonotypes detected by NoisET applied to these datasets, as well as to data from a secondary Yellow-Fever vaccination study \cite{Minervina2020}.

In the yellow fever vaccination study, TCR repertoires of three pairs of identical twins were sequenced~\cite{Pogorelyy2018}. In each donor, 600 to 1700 responding TCR clones were identified. The TCR response was highly personalized even among twins. Analyzing the clonotypes the method called responding, we were able to show that while the responding TCRs were mostly private, they could be well-predicted using a classifier based on sequence similarity. Using the a posteriori distribution, different types of dynamics were found in different TCR subsets: CD4+ cells contract faster than CD8+ cells. 

TCR cDNA-based repertoire response identified groups of CD4+ and CD8+ T cell clones that contract after recovery ($\sim 15$  days after the onset of symptoms) from a SARS-CoV-2 infection~\cite{Minervina2021}. A secondary response peak of the response was identified $\sim 40$ days after the onset of symptoms. This secondary peak was also seen in other SARS-CoV2 studies~\cite{Weiskopf2020}, however it did not correspond to known tetramer probes. Analyzing repertoire data for the same individuals taken a year and two years before the SARS-CoV2 infection, we showed that T-cell clones detected as reacting to SARS-Cov-2 were present one year before the SARS-Cov-2 infection. A network analysis revealed that these pre-existing cells that could confer immunity were specific to a SARS-CoV2 epitope with a one amino acid mutation compared to a common cold coronavirus. This observation raised the question of the correlation between the presence of cross-reactive T cells before infection and mildness of the disease. The detected reactive T-cell clones were also found in memory subpopulations at least three months after the infection.

As mentioned in section~\ref{feat_gen} the noise learning feature of NoisET has also been used to learn the natural dynamics of TCR repertoires based on gDNA and cDNA data in the absence of direct antigenic stimulation~\cite{BensoudaKoraichi2022}. This study considered the TCR$\beta$ repertoires of 9 people and showed that the dynamics of all people, regardless of age is constrained by the power law exponent of the frequency distribution. The exponent itself is given by the ratio of the deterministic turnover timescale and the stochastic noise timescale. The reproducibility of this ratio is a very strong constraint, not directly encoded by the model but learned from the statistics of the data, that implies strong amplitudes of environmental antigenic fluctuations compared to the mean fitness of lymphocyte clones. This parameter regime translates into a very susceptible dynamical system since the mean of the clone size distribution diverges. This property allows the repertoire to maintain a large number of cells, even if the source disappears or becomes very small. While the ratio is constrained, the repertoire turnover timescale shows a strong dependence on the age of the individual, with clear signatures of ageing in the physical sense: turnover timescales grow linearly with the biological age of the individual from $\sim 10$ years for 20-year olds to  $\sim 40$ years for 60-year olds. This timescale gives us an estimate of how likely we are to find clones in the repertoire after a certain number of years, depending on the person's age.
 
\section{Comparison with existing software}
The need to characterise experimental noise has been well recognized in the sequencing community. EdgeR is a package used to analyse a variety of data produced with HTS (High Throughput Sequencing) that includes read counts~\cite{edgeR}. This software has been mostly used for differential gene expression analysis, differential splicing and bisulfite sequencing. Applied to lymphocyte repertoires, the EdgeR package enables using statistical tests to identify TCR clones expanded after an acute infection. 

We compare EdgeR and NoisET detected clones assumed to respond to SARS-Cov2 antigen, based on  TCR data from Adaptive Biotechnologies (\url{https://clients.adaptivebiotech.com/pub/covid-2020}). We can validate the responding  clones using the MIRA dataset from the same group for which the reactivity to SARS-Cov2 antigen was validated experimentally~\url{https://www.ncbi.nlm.nih.gov/pmc/articles/PMC7418734/}. To count the overlap between the responding clonotypes called by each software and the TCR MIRA database, we used the AtrieGC software~(\url{https://github.com/mbensouda/NoisET_tutorial/}), which enables to rapidly compare two lists of amino-acids. In order to ascribe statistical significance to our results, we compare the numbers of overlapping TCRs called by EdgeR and NoisET  to overlapping TCRs between lists of $1000$  randomly sampled clones from the experimental samples and the MIRA list. Given the mean and standard deviation of overlapping clones, we quantify the performance of the two softwares using a $z$-score. The conclusion is drawn in a Venn diagram in Fig.\ref{fig:01}e. For this specific task of recognizing SARS-Cov2 TCR clones NoisET ($p$-value of $5.10^{-13}$)  performs similarly to edgeR ($p$-value of $2.410^{-14}$) with the benefits of better understanding of the data, better knowledge of the log-fold change statistics and the possibility to generate synthetic data. We note that the MIRA database is non exhaustive so both NoisET and   edgeR may have called truly responding SARS-Cov2 TCR clones that are not included in the MIRA database.

\section{Discussion}

High-throughput sequencing of immune repertoires is poised to revolutionize systems immunology as well as precision medicine. In particular, there is a growing interest in identifying T-cell receptors that respond to acute infections and vaccine challenges, based on experiments that probe repertoires before and after an antigenic challenge. Due to experimental and biological noise, identifying the response simply based on differences in counts before and after the challenge is not reliable. The commonly used solution is to prune these estimates using statistical tests, which are not tailored to account for these specific sources of noise. 

In our previous work, we provided a computational method that accounts for the different biological and experimental sources of noise in the clone count measurements in a Bayesian way, allowing for a more reliable detection of expanded or contracted clones. However, while the proof-of-principle algorithm explored the applicability of the method, it did not provide a user-friendly tool, which limits its wide use by the community of immunologists and clinicians. Here, we described a new computational tool, NoisET (\url{https://github.com/statbiophys/NoisET}), a python package with a command-line interface that implements the method for characterizing the noise and identifies statistically significant responding clones. The tool is applicable to datasets of diverse origin describing the clonal repertoire response to acute infections and non-stimulated long-term dynamics.

NoisET is designed as an easy-to-use package to learn the noisy statistics of sequence counts and to detect responding clones to a stimulus as reliably as possible. It captures the experimental and biological noise for both RNAseq and gDNAseq replicate technologies. Although the package has been tested on diverse datasets,  choosing and using the adequate statistical null model should be done with caution.

Among the different types of noise models offered, the negative binomial noise model is recommended to start the analysis as its running time is shorter than the two step model, while retaining the ability to account for arbitrary noise amplitudes. So far, NoisET  has been used to study the short time scale dynamics for acute infections, but could also be used to compare bulk repertoires with selected repertoires derived from functional or cultured assays \cite{Balachandran2017}. For longer time scales, the dynamics of lymphocyte populations should be modeled to best describe slow global repertoire changes that cannot be attributed to a single stimulus~\cite{Desponds2016, BensoudaKoraichi2022}.

The Bayesian approach encoded in NoisET results in a more reliable way to account for uncertainty than statistical estimates that are also less interpretable. The detection of responding clones based on the fold change of empirical abundances was not optimal without a robust interpretation of the details of the noise model. Errors in noise identification also propagate to erroneous calling of clonotypes.

From a more general perspective, NoisET and the methods behind it combine many years of the study of gene expression noise~\cite{Elowitz2002,vanOudenaarden2002,Xie2006, tanaguchi_xiescience, HornosPRE}. NoisET strongly exploits the intermittency of mRNA production and the heterogeneity of mRNA counts in individual cells. 

As we briefly discussed, NoisET has been applied to identify SARS-CoV-2-specific T-cell receptors and in the future can be used to study and understand the heterogeneity of SARS-Cov-2 vaccine response. It has potential application uncovering responding T-cell receptors to acute infections and vaccine response. 

While the method is generally applicable to T cells and B cells \cite{ourreview, Arupreview}, due to the somatic hypermutations occurring in B cells upon proliferation, care must be taken when preparing B cell data input and interpreting the model. One possibility is to collapse the sequences into lineages and consider the dynamics of a lineage in the periphery. However, while this is a reasonable first approximation, more work is needed to correctly account for the complexity of B cell repertoires. For this reason we discuss existing applications to T cells. Nevertheless, the conceptual ideas behind noise calibration as implemented in the  NoisET software apply. 

More broadly, the noise inference using the first module of NoisET has also been used to learn the natural dynamics of T cell repertoires in the absence of specific antigenic stimulation~\cite{BensoudaKoraichi2022}. For all individuals studied we found a universal contraint on the dynamics, which translated into a susceptible dynamical system that can easily maintain a large number of diverse cells. If the same type of dynamics holds for coarse grained B cell repertoires, which remains to be seen, it would point to universal laws that constrain clone size distributions and govern repertoire dynamics.

\section*{Ackowledgements}
This work was partially supported by the
European Research Council Consolidator Grant n. 724208 and ANR-19-CE45-0018 ``RESP-REP" from the Agence Nationale de la Recherche.

\bibliographystyle{pnas}

\end{document}